# Search for the Parity-Violating Effects between D- and L-alanine

Wenqing Wang    Fan Bai    Zhi Liang

*Department of Chemistry, Department of Physics, Peking University, Beijing 100871, China*

**Abstract:** The contribution of parity-violating effects to the phase transition of the D-/L-alanine crystals was confirmed by $^1$H CRAMPS solid state NMR, DC-magnetic susceptibilities and ultrasonic measurements. It was found that the spin relaxation mechanism of $\alpha$-H nucleus of D-alanine molecule is different from L-alanine and the effect is stronger than that of L-alanine. In addition, D-alanine undergoes a magnetic phase transition at a field of 1.0$T$, which is confirmed by a peak-form ultrasonic attenuation curve. DC-magnetic susceptibilities measurements of L-alanine also indicate abnormal magnetic properties, which is accompanied by a step-form ultrasonic attenuation jump and its mechanism seems different from that of D-alanine. The phase transition is considered to act as a cooperating amplification mechanism of the P-odd effects at the molecular level.

*1. Introduction.* —— The parity–violating energy difference (PVED) has become important not only to explain the selection of the L-amino acids in the origin of life, but also as a "molecular footprint" of fundamental physics[1]. Since Rein[2] and Letokhov[3] first suggested that the electroweak interaction would lift the degeneracy enantiomers predicted by normal quantum electrodynamics, a number of calculations of the PVED for small molecules have been published.[4-8] In fact, while parity violation induced by weak neutral current (WNC) is a well established fact in atoms physics[9], proof of the existence of a phenomena, which would arise from weak nuclear force in molecules, has not been reported yet. And it seems likely that the cause for the homochirality of biological molecules can be assigned to the intrinsic chirality present at the elementary particle level. [10-12]

Recently published ab initio results showed that PVED is calculated by configuration interaction singles (CIS) to be larger by one to two orders of magnitude than anticipated by the earlier single determinant excitation-restricted Hartree-Fock (RHF) methods.[4] Results, from other research groups,[5,7,8,13] concerning biologically relevant molecules, confirm larger PVED values and again indicate an energetic preference for the L-amino acids and D-sugar, which has sparked renewed interest in the experimental search parity violating (P-odd) effects in molecules.

However, Quack *et al* [4] concluded that, in addition to parity violating potentials being frequently more than an order of magnitude larger than previously calculated, earlier conclusions on the PVEDs of amino acids and sugars "are uncertain even with respect to the sign." [10,14,15] In 2000 Quack *et al* introduced a multi configuration linear reoponse (MC-LR) approach to parity violating effects for alanine in the gas and solution phase, and provided no support whatsoever for a systematic accessible conformations.[16] Meanwhile, Schwerdtfeger *et al* [17] also calculated parity-violating energy shifts for the 13 stable conformers of gaseous alanine and indicated that naturally occurring L-alanine is preferred for only seven of the investigated thirteen structures. Clearly, fundamental theoretical ambiguities currently exist in ab initio calculations both as to the magnitudes and even the directions of any PVED stabilization of these biopolymers. The question — "D- or L-alanine which is lower in energy" is reopened a discussion by theoretical physicists.[16]

A novel phase transition of D- and L-alanine crystal was first recognized by specific heat measurement with differential scanning calorimetry.[18] We select single crystal of alanine, which is a condensed state, to search for the P-odd effects between enantiomers to avoid the complicated conformational changes and solvent effects. The temperature dependent $^1$H CRAMPS solid state NMR measurements, DC-magnetic susceptibilities and ultrasonic attenuation experiments were performed.

*2. Experimental: Sample preparation/characterization.*—D-/L-alanine single crystals were characterized by elemental analysis (C, H and N) and a good agreement was shown between the theoretical and experimental data. By using X-ray diffraction crystallography at 293K, the cell dimensions of D-/ L-alanine crystals were determined as the same space group $P2_1P2_1P2_1$, orthorhombic, a = 6.0250 Å, b = 12.3310 Å, c = 5.7841 Å, V = 429.72 Å$^3$, It indicates that D-alanine and L-alanine are pure single crystals containing no crystal water. The rotation angle ζ of the D- and L- alanine solution was measured on Polarimeter PE-241 MC at 293 K with the wavelength of 589.6 nm. By using the formula of [ α ] = ζ /(L × C), the corresponding α values of D- and L-alanine were shown to be the same.

*$^1$H CRAMPS solid state NMR measurements.* $^1$H NMR multipulse spectra were run on a Varian InfinityPlus–400 spectrometer with resonance frequency 400.12 MHz. A 4mm Chemagnetics double probe was used for the variable temperature CRAMPS experiment. A BR-24 multiple sequence was employed with a π/2 pulse width of 1.6 μs and 64 scans with a 2s recycle delay to acquire CRAMPS spectra. Spin rate was 2.5 kHz and the number of scans was 64. Chemical shifts were referenced to tetramethylsilane (TMS) for $^1$H measurements.

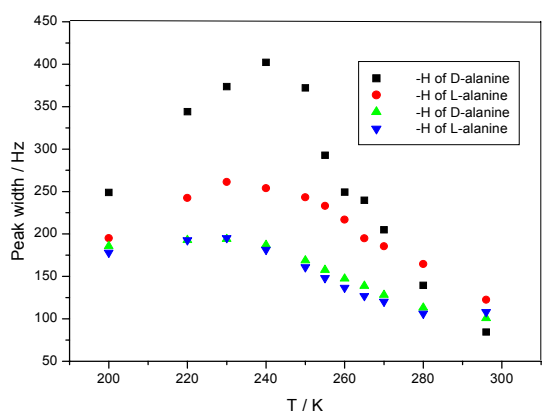

**Fig 1. Temperature dependence of peak widths of α -H, β -H of D-/L-alanine crystals**

For the sake of investigating the temperature-dependent nuclei dynamics of D- and L-alanine molecules, we measured the width of α-H and β-H peaks of D-/L-alanine. Fig.3 displays that four peak's width of α-H and β-H experience distinct maximum around 240K. So we conjecture that both D-/L-alanine may undergo a phase transition at this temperature range.

As for the L-alanine, the variation degrees of peak widths of α -H, β -H peaks make a good agreement (the maximum peak width is about two times as large as that in room temperature) in the whole process, which indicates that the temperature-dependent relaxation effects of α -H, β -H nuclei of L-alanine molecule are nearly the same in the transition process. In the case of D-alanine, the variation of α -H peak width is much fiercer than that of its enantiomer in the transition temperature range (220~250K), in addition, its transition temperature seems nearer to 240K instead of 230K. Considering the relative stability of magnetic field in the whole experimental process, these results show that, in this specific transition, the spin-spin relaxation and spin-lattice relaxation mechanism of α -H nucleus of D-alanine molecule may be different from that of L-alanine, and its relaxation effects may also stronger than its enantiomer.

*DC-Magnetic susceptibilities measurements.* Magnetic moment (*m*) and magnetic susceptibility ($\chi_\rho$) of D-/ L-alanine crystals were measured by a SQUID magnetometer (Quantum Design, MPMS-5) from 200K to 300K at a field of 1.0*T* (with differential sensitivity 1E-8 emu to 1 Tesla) in the National Laboratory for Superconductivity,

Institute Physics Chinese Academy of Sciences. Because the magnetization variation per sample was close to the baseline resolution of the magnetometer, the sample crystal was held directly in the center of the plastic straw purely by friction without using a sample holder. Crystals were weighed and determined to be 174.1mg (D-alanine), 99.5mg (L-alanine), then transferred to the straw. The signal from the plastic straw was canceled out while the temperature was measured.

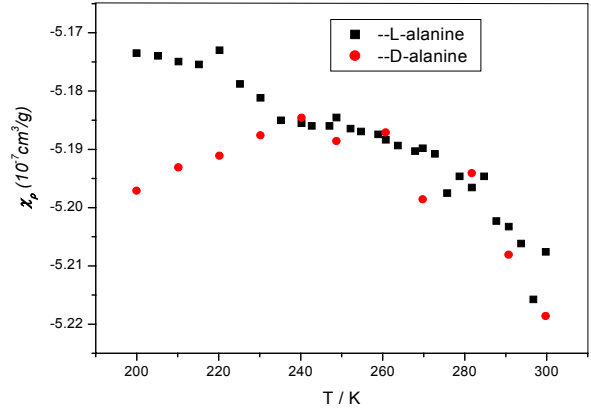

**Fig. 2. Temperature dependence of mass susceptibility at a field of 1.0 *T***

Magnetization measurements (shown in Fig.2) have shown alanine enantiomers to be dimagnetic, since they have even numbers of electrons which form closed magnetically neutral shells. It is clearly indicated that, both values and variations of magnetic susceptibilities of D-/ L- alanine enantiomers keep the same when the temperature is above 240K, and there is not distinct difference between them. It can also be found that, when temperature approach 240K from higher ones, the change of $\chi_\rho$ values becomes slow and subsiding. On the other hand, when the temperature falls below 240K, D-alanine undergoes a magnetic phase transition as the $\chi_\rho$ values showed a maximum near 240K, while $\chi_\rho$ values of L -alanine go on increasing but the rate of variation experiences an abrupt rise, which should also be looked as a magnetic phase transition and its mechanism is clearly distinct with that of D-alanine crystal. The experimental results are repeatable for the same samples after several thermal circles from 200K to 300K, and both the magnitudes of $\chi_\rho$ values and the transition temperatures of D-alanine crystal are almost the same for the different thermal circles, which indicates that these changes are completely reversible for D- and L-alanine crystals.

Meaningfully, while the temperature varies in the range of 200~240K, $\chi_\rho$ values of D-/L- alanine are nearly opposite and symmetry with the $\chi_\rho$ value ($\chi_\rho = -5.185 \times 10^{-7}$ $cm^3/g$) of D-/L- alanine at 240K taken as the base value, which may show some clues for the weak neutral current playing a contrary even parity-violating role in magnetic behavior of molecules of D-alanine and L-alanine crystals in this temperature range.

*Ultrasonic attenuation measurements.* The measurements of ultrasonic attenuation values of D-/L-alanine were performed on a computer controlled ultrasonic system (Matec model 7700). A ceramic transducer for generating a longitudinal ultrasonic wave was used. The frequency of ultrasonic wave is 5 MHz. The attenuation α value was calculated from

$$\alpha = \frac{1}{d} \ln\left[\frac{A(x_1)}{A(x_2)}\right]$$

where *d* is the thickness of the sample, *A(x₁)* and *A(x₂)* are the amplitudes of the first and second echoes respectively. The samples used for ultrasonic measurements were D- and L-alanine crystals with thickness equal to 3.00 and 2.70 mm. The transducer was fixed properly to ensure the contact with the sample and there was no air between them. The temperature of the sample was controlled by a thermocouple. In this way we can gain the ultrasonic attenuation values at various temperatures. For the sake of detecting and describing the difference of both phase transitions mechanisms more precisely, relative attenuation values curve for either enantiomer was drew, respectively: the maximum attenuation value of either enantiomer was chosen as $\alpha_{max}$, the other temperature-dependent $\alpha$ values were normalized by dividing them with $\alpha_{max}$. The experiments were performed for the cooling process from 290K to 200K.

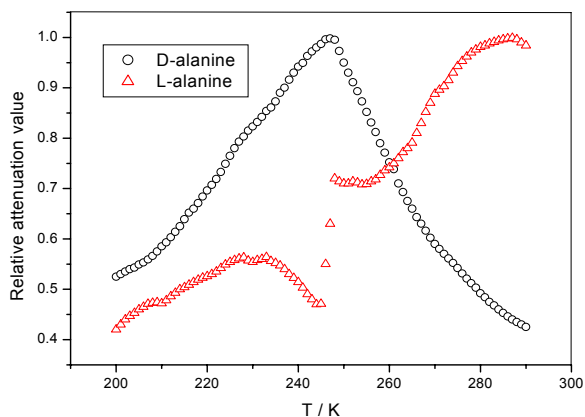

**Fig.3 The relative attenuation values of D-/L-alanine crystals vs its temperature**

The results of ultrasonic measurements are shown in Figure.3. As for the D-alanine crystal, the attenuation value changes continuously and undergoes an obvious peak around 247K, and the curve is symmetry on the whole. In the case of L-alanine crystal, a precipitous step-form drop of attenuation value was found around 247K. These ultrasonic attenuation phase transition temperatures agree with the $^1$H Solid state NMR and DC-magnetic susceptibilities measurements. It is quite clear that, for these two crystal samples, the shapes of relative attenuation curves of the enantiomers and transition mechanisms are quite different from each other. So we suppose the ultrasonic investigations demonstrate the contribution of the parity-violating WNC to the elastic relaxation effects acting in both ultrasonic transitions of D-alanine and L-alanine crystals.

Although the specific and detailed transition mechanisms which involve the parity-violating WNC is unclear, the above experimental results show that the behaviors of D-/L-alanine differ from each other in the nuclei, molecule and crystal aspects, respectively.

*3.Results and Discussions*—For the PVED scales to the fifth power of the nuclear charge of the heaviest atom in the molecule,[11] a few researchers synthesized and measured some compounds which contain heavier elements,[19,20] and for instance, very recently Lahamer *et al* [20]reported a small energy shift of the Fe(phen)$_3^{2+}$ mossbauer spectra for the two enantiomers of 0.004±0.002 mm/sec (1.9×10$^{-10}$ ev) which is larger than the theoretical calculations. The predicted ratio of the PVED is exceedingly small (10$^{-18}$) for alanine molecule without heavy element effects, and this extraordinary slight nuance is quite difficult to be detected if without any amplification mechanism for the gas and solution properties of alanine.[21] However, the influence and contribution of PVED in phase transition process of D-/L- alanine crystals, demonstrated by $^1$H solid state NMR, DC-Magnetic susceptibilities and ultrasonic measurements, is evident and significant. It has been argued that symmetry breaking could occur in condensed media in which a large number of particles can cooperate to produce a sharp transition between symmetric and asymmetric states of the samples.[22] So we conjecture that the phase transition may act as a cooperating amplification mechanism of the P-odd effects at the molecular level, of which mechanism, Salam suggested, was the Bose condensation.

In 1991 Salam[23,24] proposed that the subtle energy difference of chiral molecules induced by $Z^0$ interactions combined with Bose condensation may cause biochirality among twenty amino acids. It makes up proteins to be a consequence of second order phase transition below a critical temperature $T_c$. From Ginzburg-Landau equation, the value of temperature $T_c$ is deduced:

$$T_c = \frac{<\varphi>}{10^3} \exp[-2/g_{eff}\sigma(1-4\sin^2\theta)] \approx 2.5 \times 10^2 K$$

Here $\varphi$ field is expressed as a complex auxiliary Higgs scalar field, $g_{eff}$ is an attractive coupling constant between spin up and spin down electrons, $\theta$ is the Weinberg angle. Salam took $(1-4\sin^2\theta) \approx 1/13$, $\sigma(0)=m_z^2$ with the empirical value of the parameter $\sin^2\theta \approx 0.231$, and got $g_{eff}\sigma(0) \approx 1$.

In 1995 Figureau *et al.*[25] conducted some experiments to test the validity of Salam's prediction. After exposing both racemic and optically active cystine to temperatures ranging from 77 to 0.1K for differing time periods they observed no changes in optical rotation, thus reported failing to validate Salam's predicted phase transitions. However, after the cooling period the samples were taken out of the cryostat, heated and dissolved at room temperature in HCl solution, so their observing no optical rotation can only prove that the Salam phase transition can not be a irreversible one, but it is not a negative evidence while the transition is reversible.

The phase transitions of D-/L- alanine crystal discovered by our measurements are completely reversible and our measurements are also on-line, in addition, the phase transition temperatures (240-250K) are so considerably agreed with the theoretical estimation in Salam hypothesis, so we surmise that the phase transition may have some relation with the one Salam predicted.

Recently, the authors[26] succeed in detecting the cause of a chemical reaction using a magnetic field (in combination with unpolarized light) to the preferential production of one of the two possible enantiomers, which led to an enantiometic excess. The significance of the these findings is that now magnetochial anisotropy, apart from photochemistry with circularly polarized light and the electroweak interaction, has been a third candidate for the origin of homochirality.[27,28] In our measurements, the phenomena, that D-/L-alanine crystals in low temperature have quite different $\chi_\rho$ values and reverse variation directions, have been found to be accompanied by the contribution of the parity-violating WNC. This may provide a new sight into the relation between these two candidates for the origin of homochirality: magnetochirality and the electroweak interaction.

As we have discussed above, in this specific transition, the spin relaxation mechanism of $\alpha$-H nucleus of D-alanine molecule may be different from that of L-alanine, and its effects may also stronger than its enantiomer; in higher temperature the magnetic susceptibilities of D-/L-alanine crystals are almost the same; while the temperature below 240K, because of the P-odd amplification mechanism of the phase transition, $\chi_\rho$ values of enantiomers are quite different and vary in reverse directions; in addition, the ultrasonic transition mechanisms of alanine enantiomers are also far from each other, which is an affirmative evidence of parity violation at molecular level. Nevertheless, both questions, that the connection between the phase transitions and the one Salam predicted, and the relation between magnetochirality and the electroweak interaction, are very subtle and still open.

**Acknowledgement**

This work was supported by the grant of 863 program of China (863-103-13-06-01) .

**Reference**
1. A.J. MacDermatt, *Enantiomer*. **2000,** 5(2)**,** 153.
2. D.W. Rein, *J. Mol. Evol.* **1974,** 4,151.
3. V. S. Letokhov, *Phys. Lett.* **1975**, 53A, 275
4. A. Bakasov, T. K. Ha, and M. Quack, *J. Chem. Phys.* **1998**, 109, 7263
5. P. Lazzeretti, and R. Zanasi, *Chem. Phys. Lett.* **1997**, 279, 349
6. R. Zanasi, P. Lazzeretti, A. Ligabue, and A. Soncini, *Phys. Rev. E.* **1999**, 59, 3382
7. J. K. Laerdahl and P. Schwerdtfeger, *Phys. Rev. A.* **1999**, 60, 4439
8. J. K. Laerdahl, P. Schwerdtfeger. and H. M. Quiney, *Phys. Rev. Lett.* **2000**, 84, 3811
9. M.-A. Bouchiat, C. Bouchiat, *Rep. Prog. Phys*. **1997**, 60, 1351
10. A. Yamagata, *J. Theoret. Biol.* **1966**, 11, 495
11. D. K. Kondepudi, and G. W. Nelson, *Nature.* **1985**, 314, 438
12. M. Quack, *Chem. Phys. Lett.* **1986**, 132, 147
13. A. Bakasov, and M. Quack, *Chem. Phys. Lett.* **1999**, 303, 547
14. G. E. Tranter, *Chem. Phys. Lett.* **1985**, 115, 286
15. O. Kikuchi, and H. Wang, *Bull. Chem. Soc. Jpn.* **1990**, 63, 2751
16. R. Berger, and M. Quack, *ChemPhysChem.* **2000**, 1, 57
17. J. K. Laerdahl, R. Wesendrup, and P. Schwerdtfeger, *ChemPhysChem.* **2000**, 1, 60
18. Wenqing Wang, Fang Yi, Yongming Ni et al. *J. Biol. Phys.*, **2000**, 26**,** 51
19. A.Szabo-Nagy, and L. Keszthelyi, *Proc. Natl. Acad, Sci. USA* **1999**, 96, 4252
20. A. S. Lahamer, S. M. Mahurin, R. N. Compton, D. House, J, K. Laerdahl, M. Lein, and P. Schwerdtfeger, *Phys. Rev. Lett.* **2000**, 85, 4470,
21. M. Fujiki, *Macromol. Rapid Commun.* **2001**, 22, 669
22. M. Avalos, R. Babiano, P. Cintas, J. L. Jimenez, J. C. Palacios, *Tetrahedron: Asymmetry.* **2000**,11, 2845
23. A. Salam, *J. Mol. Evol.* **1991**, 33, 105
24. A. Salam, *Phys. Lett. B.* **1992**, 288, 153
25. A. Figureau, E. Duval, A. Boukenter, *Origins Life Evol. Biosphere* **1995**, 25, 211
26. G. L.J.A. Rikken, and E. Raupach, *Nature* **2000**, 405, 932
27. G. Wagniere, and A. Meier, *Experientia* **1983**, 39, 1090
28. L. V. Wullen, *ChemPhysChem.* **2001**, 2, 107